\documentclass[prd,aps,superscriptaddress,twocolumn,floatfix,nofootinbib]{revtex4}
\pdfoutput=1

\usepackage{amsfonts}
\usepackage{amsmath}
\usepackage{amssymb}
\usepackage{bm}
\usepackage{dcolumn}
\usepackage{graphicx}
\usepackage[latin1]{inputenc}
\usepackage{latexsym}
\usepackage{rotating}
\usepackage{hyperref}
\usepackage{graphicx}
\hypersetup{hidelinks}

\newcommand\be{\begin{equation}}
\newcommand\ba{\begin{eqnarray}}
\newcommand\ee{\end{equation}}
\newcommand\ea{\end{eqnarray}}

\begin{document}

\title{Numerical Backreaction and Finite-Time Energy Transfer in Post-Recombination Magnetogenesis from Ultralight Dark Matter}

\author{Aviv David}
\email{aviv.david@mail.mcgill.ca}
\affiliation{Department of Physics, McGill University, Montr\'{e}al,
  QC, H3A 2T8, Canada}

\author{Robert Brandenberger}
\email{rhb@physics.mcgill.ca}
\affiliation{Department of Physics, McGill University, Montr\'{e}al,
  QC, H3A 2T8, Canada}
\affiliation{Trottier Space Institute, Department of Physics, McGill
University, Montr\'{e}al, QC, H3A 2T8, Canada}

\begin{abstract}

We study gauge-field production in a post-recombination
magnetogenesis scenario driven by an ultralight pseudoscalar dark matter field $\phi$
coupled to electromagnetism.  Previous analytical work has shown that a homogeneous
oscillating $\phi$ background can excite gauge field modes through tachyonic and
narrow-band resonance channels.  Here we follow the coupled evolution of the
homogeneous $\phi$ mode and gauge field modes of both helicities in an expanding background. Our numerical work confirms the results for the spectra of produced gauge particles obtained in the previous approximate analytical treatments. In addition, our study allows us to determine when back-reaction shuts off the resonance.  We find that for parameter values for which the tachyonic instability band is open, back-reaction does not shut off the resonance until a fraction ${\cal{F}}$ of order one of the initial dark matter energy density has been transferred to photons, and this happens on a time scale which is short compared to the Hubble time.  In the parameter range in which only the narrow resonance band is open, ${\cal{F}}$ eventually reaches order one, but the time when this happens increases as the effective coupling constant decreases, and therefore ${\cal{F}}$ may remain negligible until the present time.

\end{abstract}

\maketitle

\section{Introduction}
\label{sec:intro}

It has recently been pointed \cite{BFJ} out (see also \cite{earlier}) that, in the absence of plasma effects, a coherently oscillating pseudoscalar field $\phi$ (which might constitute the Dark Matter of the Universe \footnote{The axion \cite{axion} is an example of such a field, and it was proposed as a candidate for dark matter in \cite{Fischler}.}) coupled to electromagnetism via a Chern-Simons term will induce a parametric resonance instability for long wavelength modes of the electromagnetic gauge field.  This will in turn lead to the generation of magnetic fields on cosmological scales. Depending on the values of the mass $m$ of $\phi$ and of the  coupling constant $g_{\phi \gamma}$ in the Chern-Simons term, the instability will either be of tachyonic type \cite{BFJ} or else a narrow band resonance \cite{Brahma} \footnote{The analysis has been extended to scalar field dark matter \cite{Kamali} and to vector dark matter \cite{Tatsuya}.}.  Before the time of recombination, the instability is blocked by plasma effects, but once the universe is neutral, the instability can set in \footnote{As long as the residual ionization is negligible. Whether this is the case or not is not yet resolved (see e.g. \cite{Sharma}), but based on the arguments in the Appendix of the revised version of \cite{Brahma} we believe that plasma effects can be neglected in our study.}.  Thus, the mechanism proposed in \cite{BFJ} may yield a ``late time'' scenario for the generation of cosmological magnetic fields (where the word ``late'' is meant to highlight the difference compared to  ``early time'' scenarios invoked previously (see e.g. \cite{previous}))\footnote{See also \cite{CSearly} for earlier studies of the effects of the Chern-Simons type coupling we are considering here.}.

Typically, the time scale of the instability is short compared to the time scale of the expansion of the universe, and hence the generation of the magnetic field will occur rapidly \footnote{The mathematics which describes this instability is very similar to that which is responsible for preheating after inflation \cite{TB, DK, STB, KLS1, KLS2} (see \cite{tachyonic} for a discussion of the tachyonic instability channel in this context, and \cite{ABCM, Karouby} for reviews).}.  The instability will lead to a transfer of energy from the dark matter field $\phi$ to photons.  A key question which was not answered in previous works is what fraction ${\cal{F}}$ of the energy is transferred. The answer to this question is determined by when back-reaction effects shut off the resonance.

In this paper we make a first attempt to study the termination of the resonance via back-reaction effects.  We perform a numerical simulation of the coupled pseudoscalar-gauge field equations of motion, starting from initial conditions in which $\phi$ is coherently oscillating over space, and the gauge field modes start with their vacuum initial conditions. \footnote{Note that the same physical system was studied numerically \cite{Adshead} in the context of inflationary magnetogenesis.}  As a first step, we neglect the generation of inhomogeneous modes of $\phi$, which can be justified since the analytical estimates in \cite{Brahma} indicate that the dominant back-reaction effects are due to the gauge field modes. We determine the fraction ${\cal{F}}$ of the energy which is transferred to the gauge fields.  Our main result is that both for parameter values $m$ and $g_{\phi \gamma}$ for which the tachyonic resonance channel is open and for parameter values for which the instability is of narrow-resonance type, more than half of the energy is transferred from $\phi$ to the gauge fields. The time scale for the energy transfer increases as the effective coupling constant decreases. Thus, when considering the system at a fixed time, there is a lower cutoff on the value of the coupling constant for which there is efficient energy transfer.

The confirmation from the work that the magnetogenesis mechanism suggested in \cite{BFJ, Brahma} is robust has implications not only for the mystery of the origin of magnetic fields on cosmological scales (see e.g. \cite{Durrer} for reviews) but also for the formation of super-massive black holes. As suggested in \cite{Jiao},  the mechanism may also induce a photon flux on Lyman-Werner scales of sufficient strength to prevent atomic hydrogen from forming molecular hydrogen. This would open up the Direct Black Hole Collapse channel (see e.g. \cite{Volonteri} for reviews of super-massive black hole formation).  In fact, as discussed in \cite{Ashu, Abdias}, the coupling of the cosmological magnetic field with dark matter (i.e. $\phi$) fluctuations can generate the required flux of Lyman-Werner photons. 

In the next section we discuss the physical system. Here we work in natural particle physics units. We then introduce the dimensionless variables which are used in the numerical simulations. In Section 4 we list the numerical methods and checks which we have used.  The results are presented in Section 5, and we conclude with a discussion of limitations of the current study and future prospects.

\section{Physical Setup}

We study the post-recombination axion-electrodynamics model presented in
Refs.~\cite{BFJ, Brahma}.  The dark matter component is taken to be an ultralight pseudoscalar field $\phi$ whose potential is well approximated to be quadratic over the field range relevant for the simulations.  The field is coupled to electromagnetism through the usual Chern-Simons interaction. The full Lagrangian is
\begin{equation}
\begin{split}
S=\int d^4x\sqrt{-g}\bigg[
&-\frac12 g^{\mu\nu}\partial_\mu\phi\,\partial_\nu\phi
-\frac12m^2\phi^2  \\
&-\frac14F_{\mu\nu}F^{\mu\nu}
-\frac14 g_{\phi\gamma}\phi F_{\mu\nu}\widetilde F^{\mu\nu}
\bigg] \, ,
\end{split}
\label{eq:action}
\end{equation}
where $F_{\mu \nu}$ is the field strength tensor of the usual electromagnetic field, and $\widetilde F^{\mu\nu}$ is its dual.   The sign of the last term depends on the convention used for the dual tensor and for the helicity basis.  In the simulations the relative sign is fixed by an energy-transfer check in the non-expanding limit, described in
Appendix 2.  This is a useful practical way to avoid hiding a sign
choice inside the notation for \(F\widetilde F\).

We consider a spatially flat FLRW background,
\begin{equation}
ds^2=a^2(\eta)\left(-d\eta^2+d{\bf x}^2\right),
\label{eq:flrw}
\end{equation}
where $\eta$ is conformal time,  $a(\eta)$ is the cosmological scale factor, and we work in Coulomb gauge, \(A_0=0\) and \(\nabla\cdot{\bf A}=0\).  The field $\phi$ is treated as a coherent homogeneous mode \footnote{Homogeneous initial conditions for $\phi$ can be justified by a ``misalignment'' mechanism: the value of $\phi$ which is preferred at high energy scales is not the same as the one which minimizes the energy at late times, and the preferred initial value is frozen in by Hubble friction until late times. See \cite{misalign} for a discussion of this mechanism.}. This is the leading description of the post-recombination magnetogenesis mechanism: the dark-matter field provides an oscillating background, and the gauge modes are the degrees of freedom that become unstable.

Analytical estimates in Ref.~\cite{Brahma} indicate that the \(\phi\) fluctuations generated by the gauge sector do not terminate the resonance before an order-one fraction of the energy has been transferred.  We therefore use the truncation to a homogeneous $\phi$ field, keeping only the inhomogeneous gauge field modes,  to determine the ``first nonlinear energy-transfer epoch'',  the time when backreaction becomes important.  An improved calculation of the post-transfer evolution will have to go beyond this minimal system and include inhomogeneous \(\phi\) dynamics and the subsequent plasma or magnetohydrodynamic evolution. This is being done in work which is in progress.

For a homogeneous \(\phi\), the amplitudes of the two circular polarizations of the vector
potential obey the equations
\begin{equation}
{\cal A}_{\lambda,\eta\eta}
+\left[
k^2+\lambda k g_{\phi\gamma}\phi_\eta
\right]{\cal A}_{\lambda}=0,
\qquad \lambda=\pm1 ,
\label{eq:gauge-conformal-phys}
\end{equation}
where \(k\) is the comoving wave number, and the subscript on $\phi$ indicates a derivative. Since \(\phi_\eta=a\dot\phi\),  where the overdot denotes the derivative with respect to physical time, the oscillating background periodically changes the effective frequency of the two helicities with opposite signs.  During part of each oscillation, one helicity will have a negative effective squared frequency over a range of modes, and this induces exponential growth of the mode functions. This is the tachyonic channel emphasized in Ref.~\cite{BFJ}.  The instability occurs for all infrared modes with $k < k_c$. A useful envelope estimate for the corresponding critical comoving scale $k_c$ is
\begin{equation}
k_c(\eta)\simeq g_{\phi\gamma}m\Phi(\eta)a(\eta),
\label{eq:kc}
\end{equation}
where \(\Phi\) is the slowly varying amplitude of the oscillation of $\phi$.  Note that the Floquet exponent (the rate which describes the exponential growth of the mode function) is $\mu \simeq k_c$.  If $\phi$ constitutes all of the dark matter at the time of recombination,  then
\begin{equation}
m \Phi (\eta_{\rm rec}) \, \sim \, T_{\rm rec}^2 \, 
\end{equation}
(where the subscripts refer to the values of the quantities at the time of recombination). 

The same mode equation also has a narrow parametric-resonance channel centered at
\begin{equation}
k_p\simeq \frac{a m}{2},
\label{eq:kp}
\end{equation}
as discussed in Ref.~\cite{Brahma}.  We use \(k_c\) and \(k_p\) as diagnostics in the plots of the excitation spectra.  An advantage of the numerical analysis is that we do not need to  impose a specific channel by hand; the numerics evolves all lattice modes in the sampled Fourier box, and thus directly yields the information about which instability mechanism dominates.  Note that the Floquet exponent of the instability in the narrow resonance band is
\be \label{growth}
\mu \, \simeq \, \frac{1}{4} g_{\phi \gamma} \Phi \, ,
\ee
where $\Phi$ is the amplitude of the pseudoscalar field \cite{Brahma}. The amplitude of the gauge field modes grows as $e^{\mu z}$ (where the dimensionless rescaled time $z$ is given by (\ref{eq:z-def})).

The equation for $\phi$ contains the back-reaction source
\(\langle{\bf E}\cdot{\bf B}\rangle\).  Using physical time $t$, the equation takes the form
\begin{equation}
\ddot\phi+3H\dot\phi+m^2\phi
=s\,g_{\phi\gamma}\left\langle{\bf E}\cdot{\bf B}\right\rangle_{\rm phys},
\label{eq:scalar-physical}
\end{equation}
where \(s=\pm1\) records the convention for \(F\widetilde F\).  In the linear
regime the gauge equation can be interpreted in terms of Floquet exponents, but
that description is not enough once the energy in the gauge sector becomes comparable to that in the
\(\phi\) sector.  The main numerical problem is therefore to evolve
Eqs.~\eqref{eq:gauge-conformal-phys} and \eqref{eq:scalar-physical}
self-consistently through the first strong back-reaction event.

\section{Dimensionless Variables and Equation Conventions}

The numerical analysis makes use of dimensionless variables. In this section we describe the variables which are used in the numerics and how they are related to the original physical ones.

The numerical evolution uses a rescaled dimensionless physical time variable
\begin{equation}
z=m(t-t_{\rm rec}) \, .
\label{eq:z-def}
\end{equation}
We also introduce the rescaled conformal time
\begin{equation}
\widetilde\eta=m\eta \, .
\label{eq:etatilde-def}
\end{equation}
The dimensionless wave number $K$ and Hubble parameter $h$ are
\begin{equation}
K=\frac{k}{m},\qquad h=\frac{H}{m}.
\label{eq:k-h-def}
\end{equation}
The field $\phi$ is normalized by its initial recombination-era reference amplitude
\begin{equation}
\phi = \Phi_{\rm rec}\,\varphi .
\label{eq:scalar-normalization}
\end{equation}

With \(dt=a\,d\eta\), derivatives with respect to \(z\) and \(\widetilde\eta\) are related by
\begin{equation}
\frac{d}{d\widetilde\eta}=a\frac{d}{dz}.
\label{eq:time-derivative}
\end{equation}
Thus
\begin{equation}
\varphi_{\widetilde\eta}=a\varphi_z,\qquad
\varphi_{\widetilde\eta\widetilde\eta}
=a^2\varphi_{zz}+a^2h\varphi_z .
\label{eq:scalar-derivative-conversion}
\end{equation}

In the convention used here the dimensionless effective coupling is simply
\begin{equation}
\alpha_{\rm eff}=g_{\phi\gamma}\Phi_{\rm rec}.
\label{eq:alpha-eff}
\end{equation}
There are no additional factors of \(a\) or \(m\) in this definition.  Based on the arguments of \cite{Brahma}, we expect the tachyonic resonance channel to be effective provided that the relative change in amplitude of the gauge modes during one oscillation period of $\phi$ is larger than one, and this corresponds to $\alpha_{\rm eff} \gg 1$.

In rescaled conformal time, Eq.~\eqref{eq:gauge-conformal-phys} becomes
\begin{equation}
{\cal A}_{\lambda,\widetilde\eta\widetilde\eta}
+\left[
K^2+\lambda K\alpha_{\rm eff}\varphi_{\widetilde\eta}
\right]{\cal A}_{\lambda}=0 .
\label{eq:gauge-etatilde}
\end{equation}
Using Eq.~\eqref{eq:time-derivative}, the equivalent physical-time equation is
\begin{equation}
{\cal A}_{\lambda,zz}
+h{\cal A}_{\lambda,z}
+\left[
\left(\frac{K}{a}\right)^2
+\lambda\alpha_{\rm eff}\left(\frac{K}{a}\right)\varphi_z
\right]{\cal A}_{\lambda}=0 .
\label{eq:gauge-code}
\end{equation}
The additional \(h{\cal A}_{\lambda,z}\) term is numerically tiny in the local post-recombination windows considered below, but it belongs in the physical-time equation.  Keeping it makes the physical-time evolution equation exactly consistent with the conformal-time form.

The source and energy terms in our equations involve sums over the gauge field modes. We use finite-volume canonical modes.  In a periodic box
of comoving volume \(V=L^3\),
\begin{equation}
A_i({\bf x},\eta)=\frac{1}{\sqrt V}
\sum_{{\bf k},\lambda}
{\cal A}_{\lambda,{\bf k}}(\eta)
\epsilon_i^{(\lambda)}(\hat{\bf k})e^{i{\bf k}\cdot{\bf x}},
\label{eq:mode-expansion}
\end{equation}
with the reality condition on \(A_i\) understood.  The helicity vectors satisfy
\begin{equation}
i{\bf k}\times{\boldsymbol\epsilon}^{(\lambda)}
=\lambda k\,{\boldsymbol\epsilon}^{(\lambda)}.
\label{eq:helicity-convention}
\end{equation}
The gauge amplitudes are initialized with the vacuum normalization
\({\cal A}_{\lambda,{\bf k}}\sim (2\omega_k)^{-1/2}\), and the explicit \(1/V\) factors are kept in the energy and source sums.  This is the canonical finite-volume convention; it avoids double counting the volume factor.

The evolved momentum variable is
\begin{equation}
P_{\lambda,{\bf k}}\equiv {\cal A}_{\lambda,{\bf k},z}.
\label{eq:p-def}
\end{equation}
Since \({\bf E}_{\rm phys}=-a^{-2}{\bf A}_\eta\) and
\({\bf B}_{\rm phys}=a^{-2}\nabla\times{\bf A}\), the dimensionless electric and magnetic energies scale as
\begin{equation}
\rho_E=
\frac{1}{2a^2V}\sum_{{\bf k},\lambda}
\left|P_{\lambda,{\bf k}}\right|^2,
\qquad
\rho_B=
\frac{1}{2a^4V}\sum_{{\bf k},\lambda}
K^2\left|{\cal A}_{\lambda,{\bf k}}\right|^2 ,
\label{eq:rho-eb}
\end{equation}
up to a common normalization.  In terms of the dimensionless gauge amplitudes
evolved in the code, the common factor omitted from Eq.~\eqref{eq:rho-eb} is
\[
{\cal C}_{\Phi}=\left(\frac{m}{\Phi_{\rm rec}}\right)^2 .
\]
This expresses the gauge energy in units of the initial scalar energy
\(m^2\Phi_{\rm rec}^2\).  The same factor is included in the backreaction
source.  The source entering the dimensionless version of the $\phi$ equation
is (see Appendix 1)
\begin{equation}
S_{EB}^{(z)}
=\frac{{\cal C}_{\Phi}}{a^3V}
\sum_{\bf k}K\,{\rm Re}
\left(
{\cal A}_{+,{\bf k}}^*P_{+,{\bf k}}
-{\cal A}_{-,{\bf k}}^*P_{-,{\bf k}}
\right).
\label{eq:seb}
\end{equation}
Thus \(S_{EB}^{(z)}\) is \(\langle{\bf E}\cdot{\bf B}\rangle_{\rm phys}\)
in units of \(m^2\Phi_{\rm rec}^2\), up to the convention-dependent sign.
The factor \(g_{\phi\gamma}/(m^2\Phi_{\rm rec})\) in Appendix 1 therefore
gives \(g_{\phi\gamma}\Phi_{\rm rec}S_{EB}^{(z)}
=\alpha_{\rm eff}S_{EB}^{(z)}\).  The relative sign is fixed by the energy
check in Appendix 2.  The common factor is suppressed there because it
multiplies both the gauge energy and the source and does not affect the sign
check.
With this convention the $\phi$ equation evolved in the simulations is
\begin{equation}
\varphi_{zz}+3h\varphi_z+\varphi
=\alpha_{\rm eff}S_{EB}^{(z)}.
\label{eq:scalar-code}
\end{equation}
Appendix 2 shows explicitly that, in the non-expanding limit,
Eqs.~\eqref{eq:gauge-code} and \eqref{eq:scalar-code} conserve the total
$\phi$-plus-gauge energy with this relative sign.

\section{Numerical Method and Diagnostics}

The simulations evolve the homogeneous $\phi$ mode and both gauge helicities in a
periodic Fourier box \footnote{Our code is inspired by the GABE code \cite{Giblin}.}.  The lattice is specified by the number of grid points per
direction, \(N\), and by the comoving side length, \(L\).  The fundamental and
axis Nyquist wave numbers are
\begin{equation}
K_{\rm min}=\frac{2\pi}{L},\qquad
K_{\rm Ny}=\frac{\pi N}{L}.
\label{eq:lattice-scales}
\end{equation}
Changing \(N\) mainly changes the ultraviolet reach, while changing \(L\) changes
the infrared reach.  Both checks matter near the transfer boundary, because the
onset occurs very late and small changes in the sampled modes can shift the
crossing by a few oscillations near \(z_{\rm end}\).

The runs use a matter-era local FLRW background normalized so that \(a=1\) at the
start of the post-recombination window.  Physical-unit runs use the
recombination offset
\[
z_{\rm rec}\simeq\frac{2m}{3H_{\rm rec}},
\]
so that \(h=H/m=2/[3(z+z_{\rm rec})]\).  Over the dimensionless windows evolved
here, \(h\) is small, but retaining the expansion terms keeps the normalization
and scale-factor powers consistent.  Minkowski runs are used only as sign and
energy-conservation checks.

The time integration is a staggered leapfrog scheme in the physical-time variable
\(z\).  The $\phi$ and gauge momenta are half-kicked, the fields are drifted, and
the momenta are half-kicked again using the updated background.  The baseline
production set uses
\begin{equation}
N=64,\qquad L=30,\qquad \Delta z=0.006,\qquad z_{\rm end}=1000,
\label{eq:baseline}
\end{equation}
with refined scans around the onset.  The initial conditions for $\phi$ are
\[
\varphi(0)=1,\qquad \varphi_z(0)=0 .
\]
Gauge modes begin in the canonical vacuum normalization described above.  For
each nonzero lattice mode and each helicity we draw two independent real
Gaussian variables \(X_{\lambda,{\bf k}}\) and \(Y_{\lambda,{\bf k}}\), each
with zero mean and unit variance, and set
\[
{\cal A}_{\lambda,{\bf k}}(0)
=\frac{X_{\lambda,{\bf k}}+iY_{\lambda,{\bf k}}}{\sqrt{4\omega_k}},
\qquad
P_{\lambda,{\bf k}}(0)
=-i\omega_k{\cal A}_{\lambda,{\bf k}}(0),
\]
where \(\omega_k=K/a(0)\).  This gives
\(\langle|{\cal A}_{\lambda,{\bf k}}(0)|^2\rangle=(2\omega_k)^{-1}\).
The two helicities are sampled independently, and the zero mode is set to zero.
To test robustness, we use a ten-seed ensemble
near the boundary, an \(N=96\) transfer-time curve at fixed \(L=30\), selected
\(N=128\) spot checks at \(L=30\) and \(L=60\), smaller time-step checks with
\(\Delta z=0.003\), and longer runs to \(z_{\rm end}=2000\).  The initial time is
set to \(z=0\), corresponding to recombination.  In physical-unit runs the
scale factor is evaluated using the absolute time
\(t=t_{\rm rec}+m^{-1}z\).  The time over which we evolve the system is much
smaller than a Hubble time: for \(z_{\rm end}=1000\),
\[
\Delta t=\frac{z_{\rm end}}{m}
\simeq 2.09\,m_{20}^{-1}{\rm yr}
\simeq 3.38\times10^{-6}\,m_{20}^{-1}H_{\rm rec}^{-1}.
\]

The main transfer diagnostic is the instantaneous gauge-sector fraction
\begin{equation}
F(z)=
\frac{\rho_{\rm EM}(z)}
{\rho_\varphi(z)+\rho_{\rm EM}(z)},
\qquad
\rho_{\rm EM}=\rho_E+\rho_B .
\label{eq:F}
\end{equation}
Here \(\rho_\varphi=(\varphi_z^2+\varphi^2)/2\) in the simulation
normalization, up to the small expansion corrections used internally in the diagnostics.  Two derived quantities are used throughout the results section:
\begin{equation}
F_{\rm peak}=\max_{0\le z\le z_{\rm end}}F(z),
\label{eq:Fpeak}
\end{equation}
and the first half-transfer time
\begin{equation}
z_{0.5}(\alpha_{\rm eff})
=\min\{z:F(z)=0.5\}.
\label{eq:z05}
\end{equation}
If a run does not cross \(F=0.5\) before the endpoint, it is treated as a
non-crossing point for that finite evolution time.  This distinction is essential:
after back-reaction starts, the $\phi$ and gauge sectors can exchange energy
coherently, so \(F_{\rm final}\) can be much smaller than \(F_{\rm peak}\).  The
first crossing time and the peak transfer fraction are therefore better onset
diagnostics than the endpoint value alone.

Each run writes a time series, shell-averaged spectra, peak-tracking diagnostics,
and a summary row.  The spectral figures plot \(K^3P_B(K)\) and \(K^3P_A(K)\),
normalized by the peak magnetic shell power in the displayed snapshot, and
compare the measured magnetic peak with \(K_c=k_c/m\) and \(K_p=k_p/m\).  In the
amplified band this is equivalent to plotting the excess over the initial
vacuum spectrum, while the unamplified tail remains visible many decades below
the peak.  For the near-boundary production runs used to determine the
finite-time onset, the final magnetic peaks remain well below the axis Nyquist
cutoff, with typical values
\[
K_{\rm peak}/K_{\rm Ny}\simeq0.03\text{--}0.06 .
\]
This is a useful guard against interpreting a UV-cutoff artifact as a physical
transfer threshold.

\section{Results}

The results are organized around the diagnostics which most directly display
the dynamics of the resonance and its back-reaction.  We first show representative
large-coupling histories and spectra, where the tachyonic channel is visually
clear.  We then show the corresponding small-coupling histories and spectra,
where the production is delayed and the active peak is closer to the
narrow-band scale.  We then present the transfer-time curve, the effect of
cosmological expansion on the narrow band, the fixed-time transfer fraction as
a function of $\alpha_{\rm eff}$, and the main robustness checks near the
$z_{\rm end}=1000$ boundary.  Throughout this section
\begin{equation}
F(z)=\frac{\rho_{\rm EM}}{\rho_\varphi+\rho_{\rm EM}}
\end{equation}
is the tracked gauge-sector energy fraction.

\subsection{Large-Coupling Transfer and Spectrum}

\begin{figure*}[t]
\centering
\includegraphics[width=0.70\textwidth]{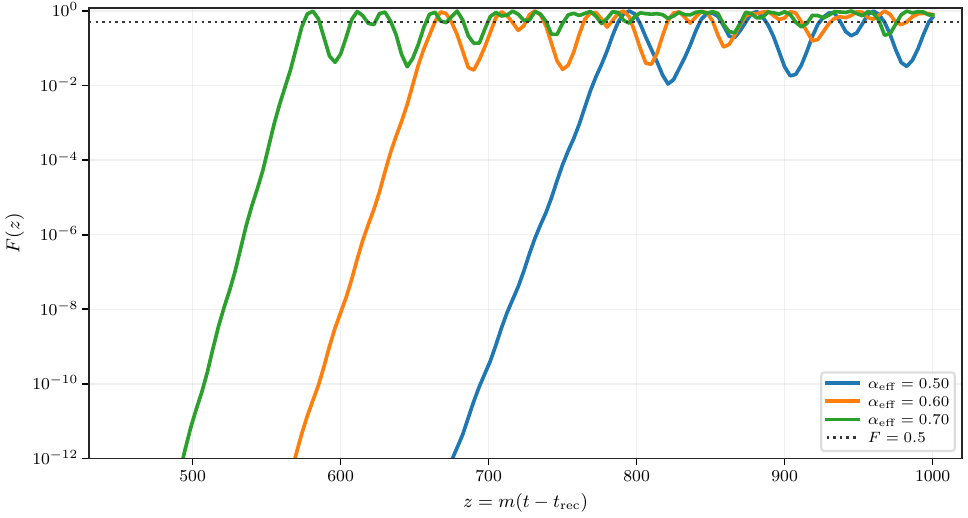}
\caption{Energy transfer $F$ as a function of rescaled time $z$ for 
representative larger values of the couplings.  The
logarithmic scale allows us to show the growth over many decades before back-reaction becomes important. The dotted line marks the half-transfer threshold $F=0.5$.}
\label{fig:large-fz}
\end{figure*}

Figure~\ref{fig:large-fz} shows representative transfer histories for
$\alpha_{\rm eff}=0.50$, $0.60$, and $0.70$.  The logarithmic scale makes both
the initial growth and the many-decade amplification visible in the same plot.
Increasing the coupling shifts the first strong transfer event earlier in $z$.
After the gauge sector becomes large, $F(z)$ does not immediately settle to a
fixed plateau.  This is why we use $z_{0.5}$ and $F_{\rm peak}$, rather than
only the endpoint value $F(z_{\rm end})$, to characterize the onset of
back-reaction.

\begin{figure*}[t]
\centering
\includegraphics[width=0.58\textwidth]{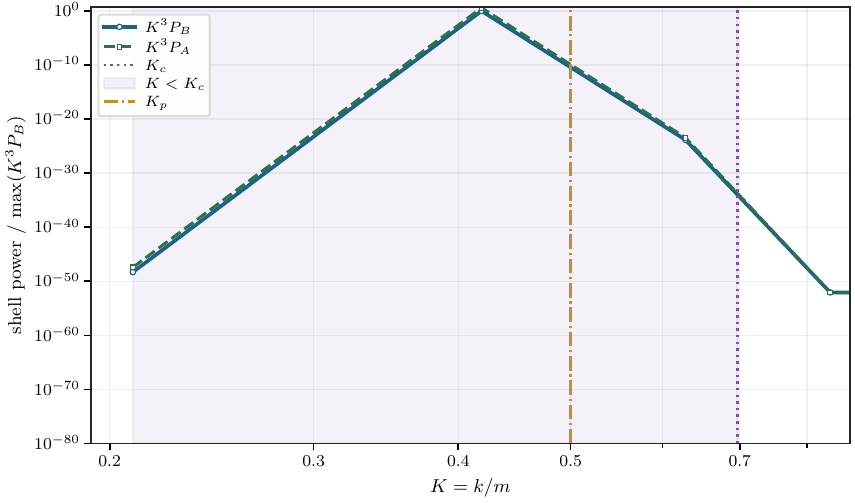}
\caption{Normalized shell-averaged $A$ and $B$ field power spectra 
near first half-transfer for
$\alpha_{\rm eff}=0.70$.  Both shell powers are divided by the peak value of
$K^3P_B$.  The vertical lines show the envelope tachyonic scale $K_c$ (dotted line) and the narrow-band scale $K_p$ (dashed line).  The dominant magnetic peak is resolved and lies inside the tachyonic envelope. The region of $K$ values where tachyonic resonance is possible is the shaded one.}
\label{fig:large-spectrum}
\end{figure*}

The corresponding normalized spectrum for $\alpha_{\rm eff}=0.70$ is shown in
Fig.~\ref{fig:large-spectrum}.  The snapshot is chosen near the first
half-transfer crossing, and the shell powers are divided by the peak magnetic
shell power to emphasize the shape and location of the amplified modes while
still showing the many-decade suppression of the neighboring shells.  The
dominant magnetic peak is resolved, lies inside the envelope tachyonic scale
$K_c$, and remains far below the Nyquist scale.  In this regime the simulation
therefore follows the behavior expected from the tachyonic analytical estimate:
one helicity is amplified efficiently during the part of the $\phi$ oscillation
for which its effective squared frequency is negative.

\subsection{Small-Coupling Transfer and Spectrum}

\begin{figure*}[t]
\centering
\includegraphics[width=0.70\textwidth]{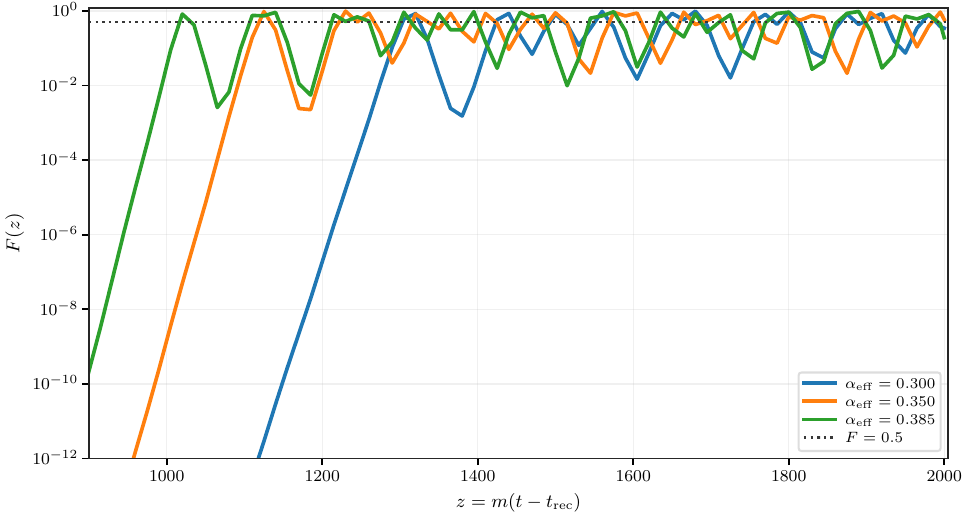}
\caption{Energy transfer $F$ as a function of $z$ for smaller 
couplings in runs extended to
$z_{\rm end}=2000$.  
%The apparent boundary in a $z_{\rm end}=1000$ scan is a
%finite-time onset boundary rather than an absolute stability threshold.  
The dotted line marks $F=0.5$.}
\label{fig:small-fz}
\end{figure*}

The smaller-coupling runs show why the result should be interpreted as a
finite-time onset boundary rather than as an absolute stability threshold (in the sense
that the energy remains in the $B$ field). After the fraction $F$ crosses the value $0.5$
for the first time, it does not remain large but oscillates back to lower values, implying
that energy flows back to $\phi$.
Figure~\ref{fig:small-fz} shows runs extended to $z_{\rm end}=2000$ for
$\alpha_{\rm eff}=0.300$, $0.350$, and $0.385$.  All three eventually reach
strong transfer, but the first half-transfer time is delayed as the coupling
decreases.  For example, $\alpha_{\rm eff}=0.300$ crosses only at
$z_{0.5}\simeq1302$.  
%If the same run were stopped at $z=1000$, it would look
%stable even though it is not stable on the longer time interval.

\begin{figure*}[t]
\centering
\includegraphics[width=0.88\textwidth]{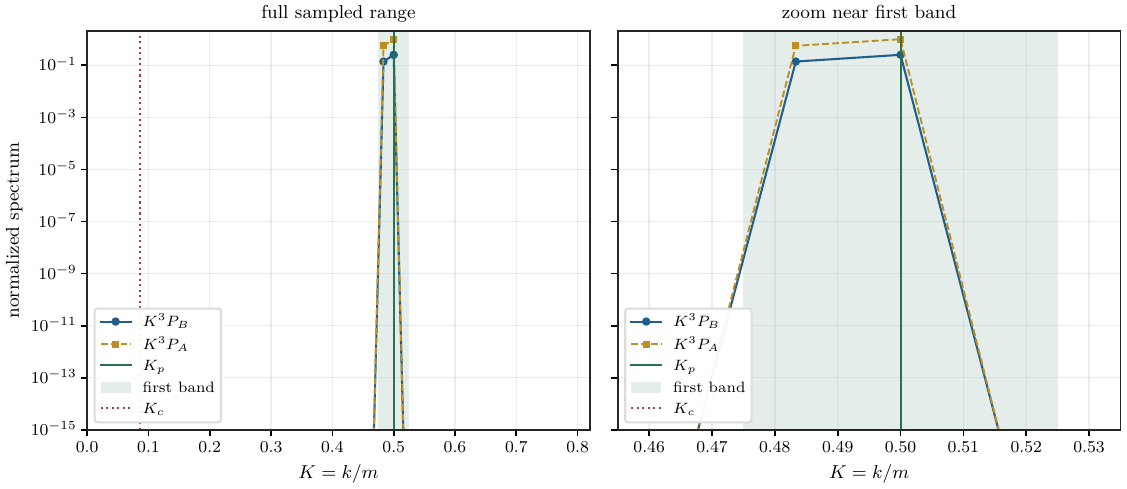}
\caption{Normalized spectrum near first half-transfer for
\(\alpha_{\rm eff}=0.1\).  The peak is at the first narrow-resonance scale
\(K_p\simeq0.5\), far outside the tachyonic envelope \(K_c\).  The left panel
shows the full sampled range, while the right panel zooms in on the narrow
band.  This is a clean example of narrow-band dominated growth. Note that in this plot the shaded region corresponds to the first narrow resonance band.}
\label{fig:small-spectrum}
\end{figure*}

The small-coupling spectrum in Fig.~\ref{fig:small-spectrum} has a different
character from the large-coupling spectrum.  With the same normalization, the
magnetic peak is outside the tachyonic envelope estimate and close to $K_p$.
We do not interpret this as a perfectly sharp separation between two isolated
mechanisms; the resonance bands are time dependent, and a finite lattice samples
only a discrete set of modes.  Nevertheless, the comparison of
Figs.~\ref{fig:large-spectrum} and \ref{fig:small-spectrum} supports the
physical interpretation of Ref.~\cite{Brahma}: the larger-coupling branch is
tachyonic dominated, while the smaller-coupling branch is more naturally
described as a delayed narrow-resonance branch.  This is a regime diagnostic
from the resolved spectra, not a separate extraction of Floquet bands.

\subsection{Transfer Time as a Function of Coupling}

\begin{figure*}[t]
\centering
\includegraphics[width=0.70\textwidth]{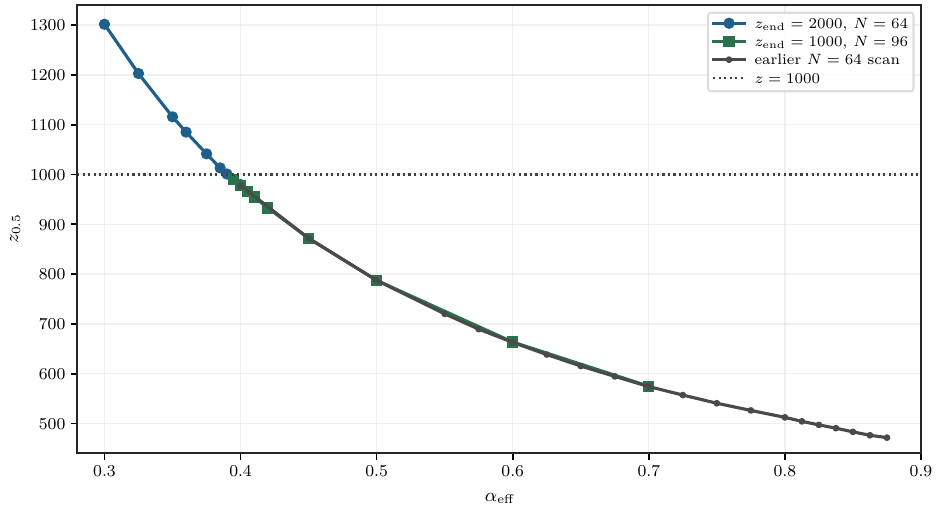}
\caption{Transfer-time (vertical axis)  as a function of the coupling constant (horizontal axis). The points are the numerical results, the solid curve is the predicted scaling from (\ref{scaling}).  Results from simulations with different resolutions are shown.}
\label{fig:alpha-scaling}
\end{figure*}

From Fig.~\ref{fig:small-fz} we see that the first half-transfer time increases as $\alpha_{\rm eff}$ decreases. This is expected from the analytical considerations. The energy density in the gauge field grows as
\be
\rho_A(z) \, \sim \, e^{2 \mu z} \rho_i \, ,
\ee
where $\rho_i$ is the initial gauge field energy density (see (\ref{growth})).  Hence, we expect the time interval to transfer half of the initial $\phi$ energy into gauge excitations to scale as
\be \label{scaling}
z_{0.5} \, \propto \, \alpha_{\rm eff}^{-1} \, .
\ee
Our numerical results for $z_{0.5}$ are shown in Figure~\ref{fig:alpha-scaling}.

\subsection{Effects of the Expansion of Space}

\begin{figure*}[t]
\centering
\includegraphics[width=0.88\textwidth]{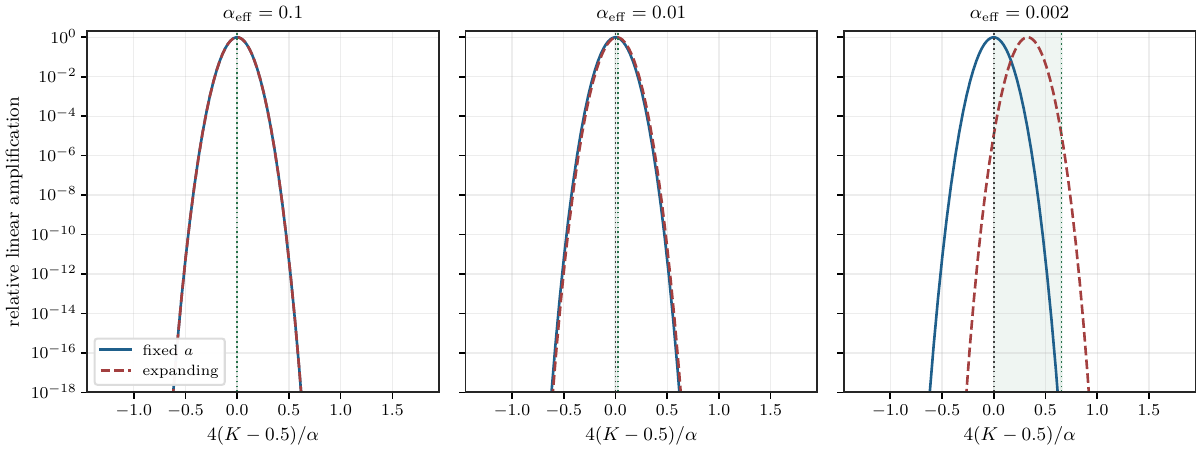}
\caption{Moving-band estimate of the relative linear amplification in the first
narrow-resonance band.  Each curve is obtained from
\(\exp[2\int\mu_K(z)\,dz]\) using the local first-band Floquet exponent and is
normalized to its own maximum; these curves are not results of the full
nonlinear simulations.  For \(\alpha_{\rm eff}=0.1\) the fixed-$a$ and
expanding estimates are essentially indistinguishable.  As the coupling is
decreased, the resonance has more time to drift, and the expansion-induced
shift toward larger comoving \(K\) becomes visible.}
\label{fig:expansion}
\end{figure*}

All results described up to now were obtained from simulations which include the expansion of space. But how important is the expansion of space? Analytical approaches typically neglect the expansion. If the time scale of the instability is much smaller than the Hubble time scale, then we expect the effects of cosmological expansion to be negligible. The time scale of the instability is inversely proportional to the Floquet exponent, i.e. inversely proportional to $\alpha_{\rm eff}$. Figure~\ref{fig:expansion} shows a moving-band estimate of the relative linear amplification of gauge modes, comparing the fixed-$a$ approximation with the expanding case. The expansion shifts the narrow resonance band toward larger comoving $K$, but the effect is only important at very small values of the effective coupling, when the instability time scale ceases to be negligible compared to the Hubble time scale.

\subsection{Fixed-Time Transfer Fraction}

\begin{figure*}[t]
\centering
\includegraphics[width=0.88\textwidth]{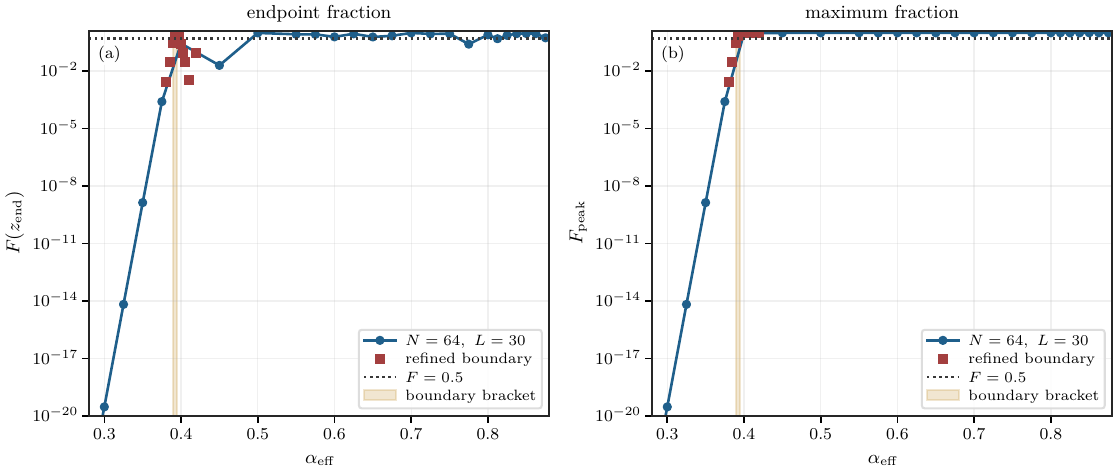}
\caption{Gauge-sector fraction as a function of $\alpha_{\rm eff}$ at a fixed time.  The
endpoint value can oscillate after transfer begins, while $F_{\rm peak}$ records
whether strong transfer occurred during the run.  The shaded band marks the
seed-stable $z_{\rm end}=1000$ onset region.}
\label{fig:falpha}
\end{figure*}

Figure~\ref{fig:falpha} shows the endpoint fraction as a function of $\alpha_{\rm eff}$.  The endpoint fraction in the left panel is useful but not sufficient by itself.  Once nonlinear transfer begins, the system can move energy back toward the $\phi$ sector before the saved endpoint.  A small endpoint value therefore does not necessarily mean that strong transfer never occurred.  The right panel shows $F_{\rm peak}$, which cleanly separates runs that entered the nonlinear transfer regime from those that did not.

The fixed-time scan in Fig.~\ref{fig:falpha} makes the finite-time
nature of the result explicit.  Below the boundary, $F_{\rm peak}$
remains small over the $z_{\rm end}=1000$ window.  Just above it, $F_{\rm peak}$
rapidly becomes order one \footnote{The ``seed-stable'' band indicates the range of values of $\alpha_{\rm eff}$ where $F$ first crosses $F=0.5$ as the random gauge-field initial conditions are varied over ten runs at each coupling.}.   The transition is sharp, but it should not be read
as an asymptotic critical coupling.  It is the coupling at which the transfer
time enters the chosen finite evolution window.

\subsection{Robustness of the Results}

\begin{figure*}[t]
\centering
\includegraphics[width=0.70\textwidth]{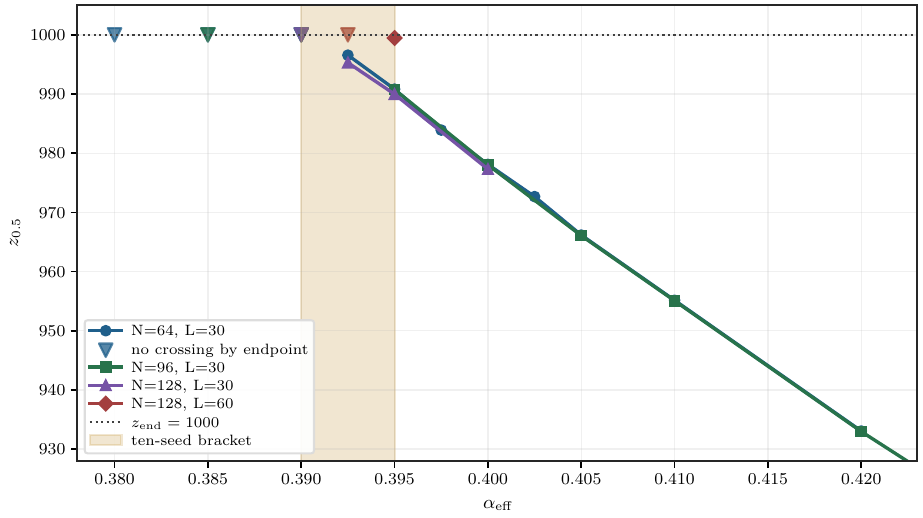}
\caption{First half-transfer time near the $z_{\rm end}=1000$ boundary.
Downward markers denote that by the endpoint of the simulation the value $F = 0.5$ has never been reached.  The shaded
band shows the seed-stable bracket
$0.390<\alpha_{\rm eff}\le0.395$. Results from simulations with different
resolutions are shown.}
\label{fig:boundary}
\end{figure*}

It is important to study the robustness of the numerical results to changes in $N$ and $L$. In Fig.~\ref{fig:boundary}, we show
the refined transfer-time boundary.  For
the baseline $N=64$, $L=30$, $\Delta z=0.006$ runs, the first single-seed
crossing is at
\begin{equation}
\alpha_{\rm eff}=0.3925,\qquad z_{0.5}=996.6 .
\end{equation}
The neighboring point $\alpha_{\rm eff}=0.390$ reaches only
\begin{equation}
F_{\rm peak}\simeq0.302
\end{equation}
by $z=1000$.  The $N=96$ curve at fixed $L=30$ agrees very well with the
$N=64$ result, and the $N=128,L=30$ spot checks support the same onset.  The
larger-box $N=128,L=60$ checks shift the latest near-threshold crossing slightly
later: $\alpha_{\rm eff}=0.3925$ does not cross by $z=1000$, while
$\alpha_{\rm eff}=0.395$ crosses very late, at $z_{0.5}\simeq999.5$.
This supports the robustness of our results.

\begin{table*}[t]
\caption{Checks of the robustness of the results.  The value of $z_{0.5}$ obtained 
does not depend sensitively on the numerical cutoffs $N$ and $L$, nor on the time discretization $\Delta z$. Results are shown for two different values of the 
effective coupling constant. 
%The agreement between time-step and
%fixed-$L$ resolution checks is very good; the larger-box point shows the
%remaining box-size sensitivity at the latest crossing.
}
\label{tab:checks}
\begin{ruledtabular}
\begin{tabular}{ccccc}
$\alpha_{\rm eff}$ & $N$ & $L$ & $\Delta z$ & $z_{0.5}$ \\
0.395 & 64  & 30 & 0.006 & 990.85 \\
0.395 & 64  & 30 & 0.003 & 990.82 \\
0.395 & 96  & 30 & 0.006 & 990.75 \\
0.395 & 128 & 30 & 0.006 & 989.97 \\
0.395 & 128 & 60 & 0.006 & 999.46 \\
0.400 & 64  & 30 & 0.006 & 978.10 \\
0.400 & 96  & 30 & 0.006 & 978.0 \\
0.400 & 128 & 30 & 0.006 & 977.30
\end{tabular}
\end{ruledtabular}
\end{table*}

Table~\ref{tab:checks} gives a compact version of the main robustness checks.
The time-step check is essentially unchanged at the precision relevant here, and
the $N=64$, $N=96$, and $N=128$ results agree at fixed $L=30$.  The larger-box
result is the only visible systematic near the endpoint.  At
$\alpha_{\rm eff}=0.395$, the fixed-$L=30$ checks agree within about one unit in
$z_{0.5}$.  We therefore quote the $z_{\rm end}=1000$ onset as a narrow
boundary region near
\begin{equation}
\alpha_{\rm eff}\simeq0.39,
\end{equation}
rather than as a single exact value.

The same onset region was also checked against a small seed ensemble.  These
seed checks are not shown separately in the main text, but they support the
same finite-time boundary region quoted above.

\section{Conclusion and Discussion}

We have numerically studied the coupled classical dynamics of a pseudoscalar
field \(\phi\) coupled to electromagnetism, under the assumption that the energy
is initially stored in a homogeneously oscillating \(\phi\) condensate, as
expected if \(\phi\) is an axion-like dark matter field.  We take vacuum initial
conditions for the gauge field modes, and we start the evolution at
recombination.  We have confirmed the results from previous approximate
analytical investigations, which indicate that for strong couplings there is a
tachyonic instability \cite{BFJ} leading to rapid energy transfer from \(\phi\)
to long-wavelength gauge modes, while for small couplings there is only a
narrow-band parametric instability \cite{Brahma}.  The numerical value for the
turnover from the tachyonic regime to the narrow-resonance regime agrees well
with expectations.  We have also computed the spectra of gauge field modes for
both large and small couplings.

The numerical analysis allows us to compute the fraction of the initial energy
in \(\phi\) which is transferred to gauge modes.  We find that both in the
tachyonic and narrow-band regimes this fraction becomes of order one before
back-reaction shuts off the resonance.  We have computed the time scale
\(\delta t\) for the energy transfer, and found that \(\delta t\) scales
inversely with the effective coupling constant.  This is in good agreement with
the approximate analytical expectation.  Once the coupling constant becomes so
small that the instability time scale approaches the Hubble time, the expansion
of space leads to a shift in the location and width of the resonance band.  It
would be interesting to study this effect in more detail.

There are two main limitations of our work.  First, we have not allowed
\(\phi\) fluctuations to develop.  As long as the potential of \(\phi\) is modeled
as quadratic, fluctuations in \(\phi\) are sourced by fluctuations of the gauge
field, and hence we expect their back-reaction effect to be suppressed at early
times.  We are currently working on an extended analysis which includes
fluctuations of \(\phi\).

Second, we have neglected effects of the
residual ionization which is present after recombination.  The reason we start
the dynamical evolution of the fields at recombination is that before then
plasma effects dominate and do not allow the resonance to develop.  At first
sight \cite{Sharma} it appears that the conductivity of the residual plasma will
impede the resonance from developing. However, it has been shown in the revised
version of Ref.~\cite{Brahma} that a resonance window persists even in the
presence of a residual plasma.  The key point is that the distance which the electrons move within a time interval corresponding to the exponential instability scale is smaller than the mean free path.  It would be of great interest to include the
effects of the residual plasma explicitly in our numerical setup.

\section*{Acknowledgement}

\noindent
A.D. acknowledges support from a Science Undergraduate Research Award (SURA) through the McGill University Faculty of Science. The research at McGill is supported in part by funds from NSERC, from the Canada Research Chair program, and from McGill University. We thank Nirmalya Brahma for extensive discussions.

\section*{Appendix 1: Physical Time and Conformal Time Conversion}
\label{app:time}

This appendix records the time-variable conversion used to check the equations.
Let
\[
z=m(t-t_{\rm rec}),\qquad \widetilde\eta=m\eta,\qquad h=\frac{H}{m},
\]
with \(dt=a\,d\eta\).  Then
\[
\frac{d}{d\widetilde\eta}=a\frac{d}{dz}.
\]
For the normalized $\phi$ field \(\varphi=\phi_{\rm phys}/\Phi_{\rm rec}\),
\[
\varphi_{\widetilde\eta}=a\varphi_z
\]
and
\[
\varphi_{\widetilde\eta\widetilde\eta}
=\frac{d}{d\widetilde\eta}(a\varphi_z)
=a\frac{d}{dz}(a\varphi_z)
=a^2\varphi_{zz}+a^2h\varphi_z .
\]
Starting from the physical-time $\phi$ equation
\[
\varphi_{zz}+3h\varphi_z+\varphi
=s\,\frac{g_{\phi\gamma}}{m^2\Phi_{\rm rec}}
\langle{\bf E}\cdot{\bf B}\rangle_{\rm phys},
\]
and multiplying by \(a^2\), one obtains
\[
\varphi_{\widetilde\eta\widetilde\eta}
+2a^2h\varphi_z+a^2\varphi
=s\,\frac{g_{\phi\gamma}a^2}{m^2\Phi_{\rm rec}}
\langle{\bf E}\cdot{\bf B}\rangle_{\rm phys}.
\]
Using \(\varphi_{\widetilde\eta}=a\varphi_z\) and
\({\cal H}/m=ah\), this can also be written as
\[
\varphi_{\widetilde\eta\widetilde\eta}
+2\frac{{\cal H}}{m}\varphi_{\widetilde\eta}
+a^2\varphi
=s\,\frac{g_{\phi\gamma}a^2}{m^2\Phi_{\rm rec}}
\langle{\bf E}\cdot{\bf B}\rangle_{\rm phys}.
\]

The gauge equation in the rescaled conformal-time convention is
\[
{\cal A}_{\lambda,\widetilde\eta\widetilde\eta}
+\left[
K^2+\lambda K\alpha_{\rm eff}\varphi_{\widetilde\eta}
\right]{\cal A}_{\lambda}=0 .
\]
Using
\[
{\cal A}_{\widetilde\eta}=a{\cal A}_z,\qquad
{\cal A}_{\widetilde\eta\widetilde\eta}
=a^2{\cal A}_{zz}+a^2h{\cal A}_z,
\]
and \(\varphi_{\widetilde\eta}=a\varphi_z\), division by \(a^2\) gives
\[
{\cal A}_{\lambda,zz}
+h{\cal A}_{\lambda,z}
+\left[
\left(\frac{K}{a}\right)^2
+\lambda\alpha_{\rm eff}\left(\frac{K}{a}\right)\varphi_z
\right]{\cal A}_{\lambda}=0 .
\]
This is the physical-time form used in the numerical evolution.

\section*{Appendix 2: Source Normalization and Energy Check}
\label{app:source}

The finite-volume helicity expansion used in the simulations is
\[
A_i({\bf x},\eta)
=\frac{1}{\sqrt V}
\sum_{{\bf k},\lambda}
{\cal A}_{\lambda,{\bf k}}(\eta)
\epsilon_i^{(\lambda)}(\hat{\bf k})e^{i{\bf k}\cdot{\bf x}},
\qquad V=L^3 ,
\]
with
\[
i{\bf k}\times{\boldsymbol\epsilon}^{(\lambda)}
=\lambda k\,{\boldsymbol\epsilon}^{(\lambda)} .
\]
The physical fields are
\[
{\bf E}_{\rm phys}=-\frac{1}{a^2}{\bf A}_\eta,\qquad
{\bf B}_{\rm phys}=\frac{1}{a^2}\nabla\times{\bf A}.
\]
Since \(P_{\lambda,{\bf k}}={\cal A}_{\lambda,{\bf k},z}\) and
\({\cal A}_\eta=maP\), the source is
\[
\begin{split}
\left\langle{\bf E}_{\rm phys}\cdot{\bf B}_{\rm phys}\right\rangle
\ =&-\frac{m^2}{a^3V}
\sum_{\bf k}K\,{\rm Re}
\left(
{\cal A}_{+,{\bf k}}^*P_{+,{\bf k}}
\right.\\
&\left.
-{\cal A}_{-,{\bf k}}^*P_{-,{\bf k}}
\right).
\end{split}
\]
It is therefore convenient to define
\[
S_{EB}^{(z)}
=\frac{1}{a^3V}
\sum_{\bf k}K\,{\rm Re}
\left(
{\cal A}_{+,{\bf k}}^*P_{+,{\bf k}}
-{\cal A}_{-,{\bf k}}^*P_{-,{\bf k}}
\right),
\]
so that
\[
\langle{\bf E}\cdot{\bf B}\rangle_{\rm phys}=-m^2S_{EB}^{(z)}
\]
for the helicity convention above.  A different convention for either the
electric field sign or the helicity basis flips this overall sign, but not the
factor of \(1/V\) or the scale-factor power \(a^{-3}\).

The relative sign in the $\phi$ equation is fixed by the non-expanding limit.
Set \(a=1\) and \(h=0\).  The gauge equations are
\[
{\cal A}_{+,zz}+\left(K^2+\alpha_{\rm eff}K\varphi_z\right){\cal A}_+=0,
\]
\[
{\cal A}_{-,zz}+\left(K^2-\alpha_{\rm eff}K\varphi_z\right){\cal A}_-=0.
\]
Using
\[
\rho_{\rm EM}
=\frac{1}{2V}\sum_{{\bf k},\lambda}
\left(|P_{\lambda,{\bf k}}|^2+K^2|{\cal A}_{\lambda,{\bf k}}|^2\right),
\]
one finds
\[
\frac{d\rho_{\rm EM}}{dz}
=-\alpha_{\rm eff}\varphi_z S_{EB}^{(z)}.
\]
If the $\phi$ equation is
\[
\varphi_{zz}+\varphi=\alpha_{\rm eff}S_{EB}^{(z)},
\]
then
\[
\rho_\varphi=\frac12(\varphi_z^2+\varphi^2)
\]
satisfies
\[
\frac{d\rho_\varphi}{dz}
=+\alpha_{\rm eff}\varphi_z S_{EB}^{(z)}.
\]
Thus
\[
\frac{d}{dz}\left(\rho_\varphi+\rho_{\rm EM}\right)=0
\]
in Minkowski space.  This check fixes the relative signs used in the simulations.

\section*{Numerical Campaigns and Outputs}
\label{app:campaigns}

The manuscript figures are generated directly from saved numerical outputs.  The
main production campaigns are:

\begin{itemize}
\renewcommand{\labelitemi}{$\bullet$}
\item Refined boundary scan:
$\alpha_{\rm eff}=0.380,\ldots,0.420$, $N=64$, $L=30$,
$z_{\rm end}=1000$;
\item Seed ensemble:
10 seeds at $\alpha_{\rm eff}=0.385$, $0.390$, $0.395$, $0.400$,
and $0.405$;
\item Higher-resolution curve:
$N=96$, $L=30$, $z_{\rm end}=1000$;
\item $N=128$ spot checks:
$L=30$ and $L=60$ near the boundary;
\item Long-time runs:
$N=64$, $L=30$, $z_{\rm end}=2000$.
\end{itemize}

Every run stores a time series, peak-tracking diagnostics, shell spectra, a
configuration record, and a compact summary row.  A plotting script reads these
files and writes both the manuscript figures and the compact tables used for
cross-checks.  This procedure keeps the plotted quantities tied directly to the
saved numerical outputs rather than to hand-entered values.

\section*{Appendix 3: Implications for Cosmological Magnetic Field Generation}
\label{app:phenom}

\begin{figure*}[t]
\centering
\includegraphics[width=0.88\textwidth]{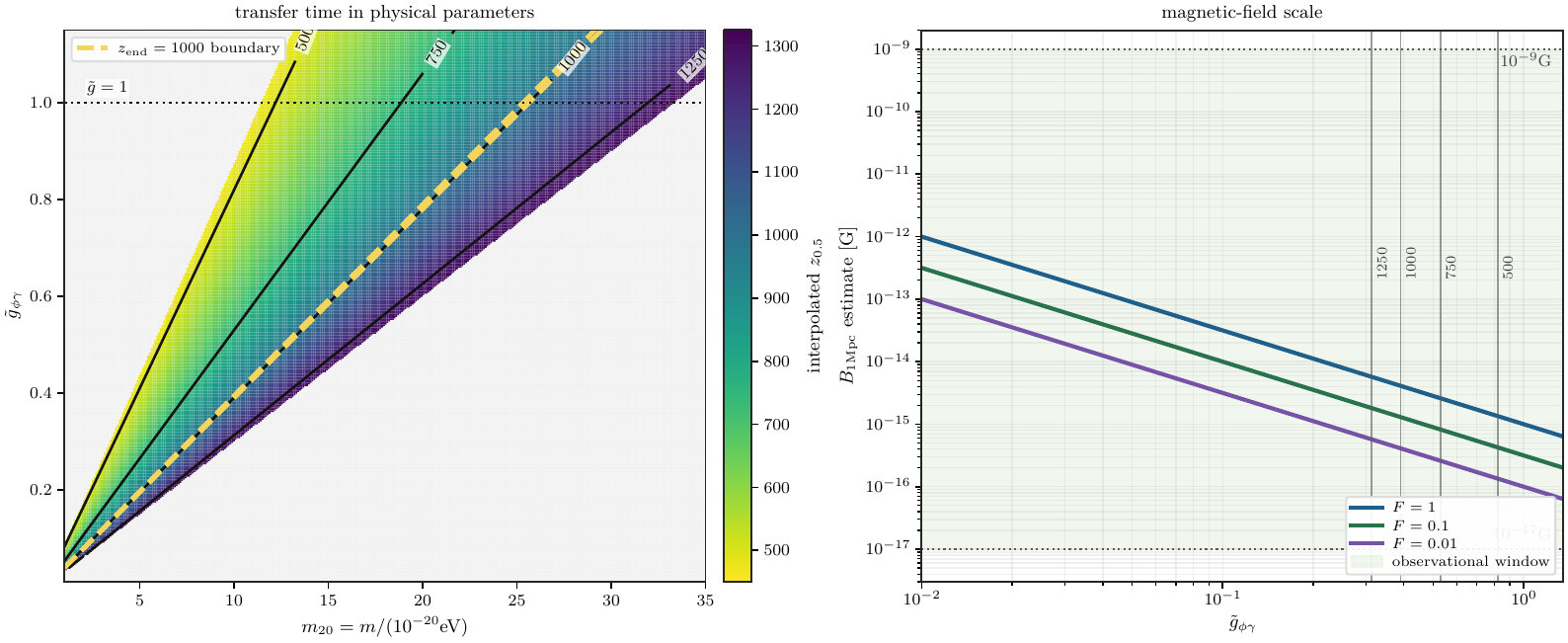}
\caption{
%Orientation map corresponding to the phenomenological conversion in
%Appendix 3.  
The left panel shows the numerically obtained
interpolated values of the transfer time
$z_{0.5}(\alpha_{\rm eff})$ (color coding) as a function of the two model parameters
$m_{20}$ and $\widetilde g_{\phi\gamma}$ using
$\widetilde g_{\phi\gamma}=0.1\alpha_{\rm eff}m_{20}$.  The solid and yellow dashed
lines correspond to fixed values of $z_{0.5}$.  The right panel depicts the
value of the produced cosmological magnetic-field scale (vertical axis) from the equation
$B_{1{\rm Mpc}}\sim10^{-15}{\rm G}\,
\widetilde g_{\phi\gamma}^{-3/2}F^{1/2}$ as a function of $\widetilde g_{\phi\gamma}$
(horizontal axis) for various values of the
energy transfer fraction $F$.  The shaded region shows the range of
$B$ values between the observational
lower bound from blazar observations \cite{obs} and the upper bound
coming from CMB (Cosmic Microwave Background) constraints \cite{Durrer}.
This figure should be viewed as a rough
scaling estimate for the magnetic field which can be generated,  and is not a 
substitute for a full late-time MHD calculation.}
\label{fig:phenom}
\end{figure*}

Let us consider a couple of phenomenological implications (the results are 
displayed in Figure (\ref{fig:phenom})), using the same
scaling variables as in Refs.~\cite{BFJ, Brahma},
\[
m=m_{20}\,10^{-20}{\rm eV},\qquad
g_{\phi\gamma}=\widetilde g_{\phi\gamma}\,10^{-10}{\rm GeV}^{-1}.
\]
The recombination-normalized \(\phi\)-field amplitude gives the approximate mapping
\[
\alpha_{\rm eff}\simeq
10\,\frac{\widetilde g_{\phi\gamma}}{m_{20}},
\]
or
\[
\widetilde g_{\phi\gamma}\simeq0.1\,\alpha_{\rm eff}m_{20}.
\]
Interpolating the measured transfer-time curve gives the benchmark values
\[
\begin{array}{c|cccc}
z_{0.5} & 500 & 750 & 1000 & 1250\\ \hline
\alpha_{\rm eff} & 0.820 & 0.530 & 0.390 & 0.313
\end{array}
\]
over the range where the interpolation is supported by the numerical runs.

For orientation we combine this with the magnetic-field estimate of
Ref.~\cite{BFJ},
\[
B_{1{\rm Mpc}}\sim10^{-15}{\rm G}\,
\widetilde g_{\phi\gamma}^{-3/2}F^{1/2}.
\]
This estimate is not used as a replacement for evolving the measured magnetic
spectrum.  It is a quick way to see where the numerically measured order-one
transfer branch lies relative to the usual intergalactic-field scale (see Figure \ref{fig:phenom}).  A final
phenomenological prediction should also include the detailed magnetic spectrum,
the precise duration of the post-recombination production window, and later MHD
evolution.

\end{document}